\documentclass[12pt,preprint]{aastex}

\slugcomment{Not to appear in Nonlearned J., 45.}

\shorttitle{Constraining Primordial Magnetic Field from CMB}
\shortauthors{Yamazaki et al.}
\begin{document}
\title{Constraining Primordial Magnetic Field from CMB Anisotropies at Higher Multipoles}
\author{D. G. Yamazaki\altaffilmark{1,2}, K. Ichiki\altaffilmark{1,2} and T. Kajino\altaffilmark{2,1}}
\email{yamazaki@th.nao.ac.jp}
\altaffiltext{1}{Department of Astronomy, Graduate School of Science, University of Tokyo, 7-3-1 Hongo, Bunkyo-ku, Tokyo, 113-0033, Japan}
\altaffiltext{2}{Division of Theoretical Astronomy, National Astronomical Observatory Japan, 2-21-1, Osawa, Mitaka, Tokyo, 181-8588, Japan}

\begin{abstract}
The cosmological magnetic field is one of the important physical quantities
 which affect strongly the cosmic microwave background (CMB) power spectrum.
Recent CMB observations have been extended to higher multipoles $l\gtrsim$1000, and they resultantly exhibit an excess power than the standard model prediction in cosmological theory which best fits the Wilkinson Microwave Anisotropy
Probe (WMAP) data at lower multipoles $l\lesssim$900.  
We calculate the CMB temperature anisotropies generated by the power-law magnetic field at the last scattering surface (LSS) in order to remove the tension between theory and observation at higher multipoles and also place an upper limit on primordial magnetic field. 
In our present calculation we take account of the effect of ionization ratio exactly without approximation.
This effect is very crucial to precisely estimate the effect of the magnetic field on CMB power spectrum.
We consider both effects of the scalar and vector modes of magnetic field on the CMB anisotropies, where current data are known to be insensitive to the tensor mode which we ignore in the present study.
In order to constrain the primordial magnetic field, we evaluate likelihood function of the WMAP data in a wide range of parameters of the magnetic field strength $|\mathbf{B}|_\lambda$ and the power-law spectral index $n_B$,  along with six cosmological parameters in flat Universe models, using the technique of the Markov Chain Monte Carlo(MCMC) method.
We find that the upper limit at $2\sigma$ C.L. turns out to be $|\mathbf{B}_\lambda|\lesssim 3.9 $nG at 1 Mpc for any $n_B$ values, which is obtained by comparing the calculated result including the Sunyaev-Zeldovich(SZ) effect with recent WMAP data of the CMB anisotropies.
\end{abstract}

\keywords{cosmic microwave background --- methods: numerical --- magnetic field} 
\section{Introduction}
Temperature and polarization anisotropies in CMB
 provide very precise information of the physical processes in the
early Universe.
However, recent new CMB data sets from the WMAP (Bennett, et al. 2003), 
the Arcminute Cosmology Bolometer Array Receiver (ACBAR; Kuo et al. 2004), 
and the Cosmic Background Imager (CBI; Mason et al. 2003) 
have indicated a potential discrepancy between theory and observation at higher
multipoles $l \ge 900$.
The best-fit cosmological model to the WMAP data predicts the power spectrum 
which shows appreciable departure from those observed in balloon and
interferometer experiments. This discrepancy cannot
be explained by taking account of tuned standard cosmological parameters.

One possible interpretation of an excess power at high multipoles
is a manifestation of the re-scattering of CMB photons by hot electrons
in clusters known as the SZ effect (Sunyaev and
Zeldovich 1980). Although there
can be no doubt that some contribution from the SZ effect exists in
the observed angular power spectrum, it has not yet been established conclusively that this is the only possible
interpretation (Aghanim et al. 2001) of the small scale power. Indeed,
the best value of the matter fluctuation amplitude to fit
the excess power at high multipoles is near the upper end of the range of 
the values deduced by the other independent method (Bond et al., 2002; 
Komatsu and Seljak 2002). Toward the solution of this bothersome
problem, it has recently been reported that the bump feature in the primordial
spectrum gives  the better explanation for both CMB and the matter power
spectra at small scales (Mathews et al., 2004).

The inhomogeneous cosmological magnetic field generated before CMB last
scattering epoch is also a possible candidate to reconcile the tension in such
higher multipoles. It excites an Alfven-wave mode in the baryon-photon
plasma in the early Universe and induces small rotational velocity 
perturbations. Since the mode can survive on scales below those of the Silk
damping during recombination (Jedamzik et al., 1998; Subramanian et al.,
1998), it could be a new source of the CMB anisotropies in such small
scales. 
Analytic expressions of temperature and polarization angular power
spectra in rather larger angular scales ($l \le 500$) based
upon the thin LSS approximation were derived for
both vector and tensor modes (Mack et al., 2002). Subramanian and Barrow
(2002) considered the vector perturbations in the opposite limit of
smaller angular scales. 
Too strong magnetic field in the early Universe, however, has a
possibility of conflicting with the cosmological observations currently
available. The  combination of those studies and current observations
places a rough bound on the strength of primordial magnetic field 
$|\mathbf{B}_{\lambda}| < (1.0-10)$nG.

In order to compare the theoretical CMB anisotropies calculated by including a primordial
magnetic field,  with observations much more precisely, we need to perform
numerical calculations of the fully linearized equations with the use of realistic recombination history of the Universe. In particular, we first need to develop a numerical method to predict the theoretical spectrum for intermediate angular scales, in which the analytic approximation becomes inappropriate. The numerical study of only the scalar mode was introduced by Koh and Lee (2000), and they discussed the sensitivity of CMB to the primordial magnetic field strength. Lewis (2004) has recently shown the characteristic of a primordial magnetic field effect on CMB in vector and tensor modes, numerically.
However, there is no systematic study to put constraint on parameters of primordial magnetic field by taking  account of both scalar and vector modes simultaneously.

The purpose of this Letter is to
explore more completely the effects of a primordial magnetic field on the CMB anisotropies, 
and to place a new limit on the field strength together with many other cosmological parameters using the MCMC method in the analysis numerically. 
The present calculation differs from all previous discussions in the following important ways. First, we take account of the effect of ionization ratio exactly without approximation.
Second, we consider the effect of scalar mode of magnetic field more precisely than the analysis of Koh and Lee (2000) by introducing the cutoff wave number physically consistent with the magneto-hydrodynamic (MHD) study of damping Alfven wave (Jedamzik et al., 1998; Mack et al., 2002). 
The effect of scalar mode of primordial magnetic field changes the CMB power spectrum significantly at larger scales for $l<900$ (Koh and Lee, 2000), while the effect of vector mode, which we also consider in the present study, dominates at smaller scales.
These effects are very crucial to quantitatively estimate the effects of magnetic field on the CMB power spectrum over a wide range of multipoles $l$ for various scales.
Third, it is for the first time to discuss the primordial magnetic field 
as a possible solution to remove the tension between theory and 
observation which exhibits in smaller angular scales of the CMB anisotropies.

\section{Primordial Stochastic Magnetic Field and Baryon-Photon Fluid}

Before recombination, Thomson scattering between photons and
electrons and Coulomb interactions between electrons and
baryons were sufficiently rapid that the photon-baryon system
behaves as a single tightly coupled fluid.
Since the trajectory of plasma particles is bent by Lorentz force in
magnetic field, photons are indirectly influenced by the magnetic field
through Thomson scattering. 
 
Let us consider the primordial magnetic field created at some moment during the radiation-dominated epoch. 
The energy density of the magnetic field is treated as a first order
perturbation in a flat Friedmann-Robertson-Walker (FRW) background
cosmology. 
Within the linear approximation, the magnetic field evolves as a stiff
source, and therefore we can discard all back reactions from the MHD fluid onto the field itself.
Also we assume that the conductivity of the primordial plasma in the
early Universe is infinite before decoupling, 
which is a very good approximation at the epochs which we are interested in.
On this assumption we can decouple the time evolution of the magnetic
field from its spatial dependence $\mathbf{B}(\eta,\mathbf{x}) =
\mathbf{B_0}(\mathbf{x})/a^2$ for very large scale, where $\eta$ is the
conformal time, $\mathbf{x}$ is the spatial coordinate, and $a$ is the
scale factor.
We also assume that a background primordial magnetic field
$\mathbf{B}_0$ is statistically homogeneous, isotropic and random. For
such a class of magnetic field, the power spectrum can be taken as a
power-law $P(k)\propto k^n$\cite{m1}, where $k$ is the wave number in
the Fourier space, and $n_B$ is the power-law spectral index of the
primordial magnetic field which can be either negative or positive.

Evaluating the two-point correlation function of the electromagnetic
stress-energy tensor of the vector mode, we can obtain the isotropic spectrum
\begin{eqnarray}
|\Pi^{(1)}(\mathbf{k})|^2\simeq
\frac{1}{4(2n_B+3)}\left[\frac{(2\pi)^{n_B+3}B^2_\lambda}{2\Gamma
\left(\frac{n_B+3}{2}\right)k^{n_B+3}_\lambda}\right]^2 
\times\left(k^{2n_B+3}_D+\frac{n_B}{n_B+3}k^{2n_B+3}\right),\label{eq:Pi}
\end{eqnarray}
for $k<k_D$ to a very good approximation (Mack et al. 2002)
\footnote{This has similar form to the one used in Mack et
al. (2002) which treats two kinds of Fourier transforms of different
normalizations. We here adopt the same normalization to the Fourier
transform systematically in order to remove uncertainty from numerical
calculations, as was pointed out by Lewis (2004)},  
where the vector mode Lorentz force, $L^{(1)}(\mathbf{k})$, is given by
$L^{(1)}(\mathbf{k})=k\Pi^{(1)}(\mathbf{k})$, $B_\lambda=|\mathbf{B}_\lambda|$ is the
magnetic comoving mean-field amplitude obtained by smoothing over a
Gaussian sphere of comoving radius $\lambda$, and $k_\lambda =2\pi/\lambda$ ($\lambda=1$Mpc in this paper).
In this equation $k_D$ is the cutoff wave number in the magnetic power
spectrum defined by 
\begin{eqnarray} 
k_D\simeq (1.7\times 10^2)^{\frac{2}{n_B+5}}\left(\frac{B_\lambda}{10^{-9}\mbox{G}}\right)^{-\frac{2}{n_B+5}}
\times\left(\frac{k_\lambda}{1\mbox{Mpc}^{-1}}\right)^{\frac{n_B+3}{n_B+5}}h^{\frac{1}{n_B+5}},\label{eq:kd}
\end{eqnarray}
where $h$ is the Hubble parameter in units of 100km/s/Mpc (Jedamzik et al. 1998; Mack et al. 2002).
Since the magnetic field source term 
$\Pi^{(1)}({\mathbf k})$ depends
on the magnetic field quadratically, the explicit time dependence 
of the magnetic stress is given by 
$\Pi^{(1)}(\eta,{\mathbf k})=\Pi^{(1)}({\mathbf k})/a^4$.
 Also the electromagnetic
stress-energy tensor of the scalar mode is $\Pi^{(0)}\simeq 2\Pi^{(1)}$(Koh and Lee, 2000; Mack et al. 2002).
In the previous studies the ionization ratio in the early Universe was
assumed to be a step function of time, i.e. $x_e=1$ before LSS and
$x_e=0$ after that. However, we have to incorporate a correct
recombination history in order to obtain more accurate theoretical
result. We accomplish this by using a numerical program 
RECFAST (Seager et al. 1999) in open use. 
Thus we take account of the correct ionization ratio $x_e(\eta)$, and we
can now rewrite  $L^{(1)}(\eta,
\mathbf{k})=kx_e(\eta)\Pi^{(1)}(\mathbf{k})$.

Combining Einstein equations with the fluid equations  (Ma and
Bertschinger 1995, Hu and White 1997), 
we obtain evolution equations of scalar and vector perturbations.
We evaluated the likelihood functions of WMAP data (Verde et al., 2003)
in a wide range of parameters of stochastic magnetic field, $B_\lambda$
and $n_B$, with other cosmological parameters, $h, \Omega_b h^2,\Omega_m
h^2, n_s, A_s$, and $\tau$ in flat Universe models, where  $\Omega_b h^2$ and $\Omega_m h^2$ are
the baryon and cold dark matter densities, $n_s$ and $A_s$ are the
spectral index and the amplitude of primordial scalar fluctuation,
and $\tau$ is the optical depth. To explore the parameter space, we make
use of the Markov chain technique (Lewis 2002).
We also take account of the SZ effect in our analysis. For that, we
follow an estimate of Komatsu and Seljak, with $\sigma_8=0.9$ (Spergel
et at. 2003; Komatsu and Seljak 2002).
Note that, we consider linear perturbations in flat FRW Universe, 
but we do not consider the effect of gravity wave damping (Caprini and Durrer, 2002) which would be relevant for the magnetic fields on super-horizon scales if they were generated before the epoch of BBN. Although this might alter the original spectrum of primordial magnetic field by the time of LSS in blue power-spectral indices $-2 < n$ (Caprini and Durrer, 2002), one can still allowed to represent the modified spectrum in running power-law index.

\section{Result and Discussions}

Our study confirms that the effect of not only the vector mode but
the scalar mode magnetic field plays the important role 
in order to solve the potential discrepancy of the CMB anisotropies 
between theory and observation at higher multipoles.

\begin{figure}
\epsscale{0.8}
\plotone{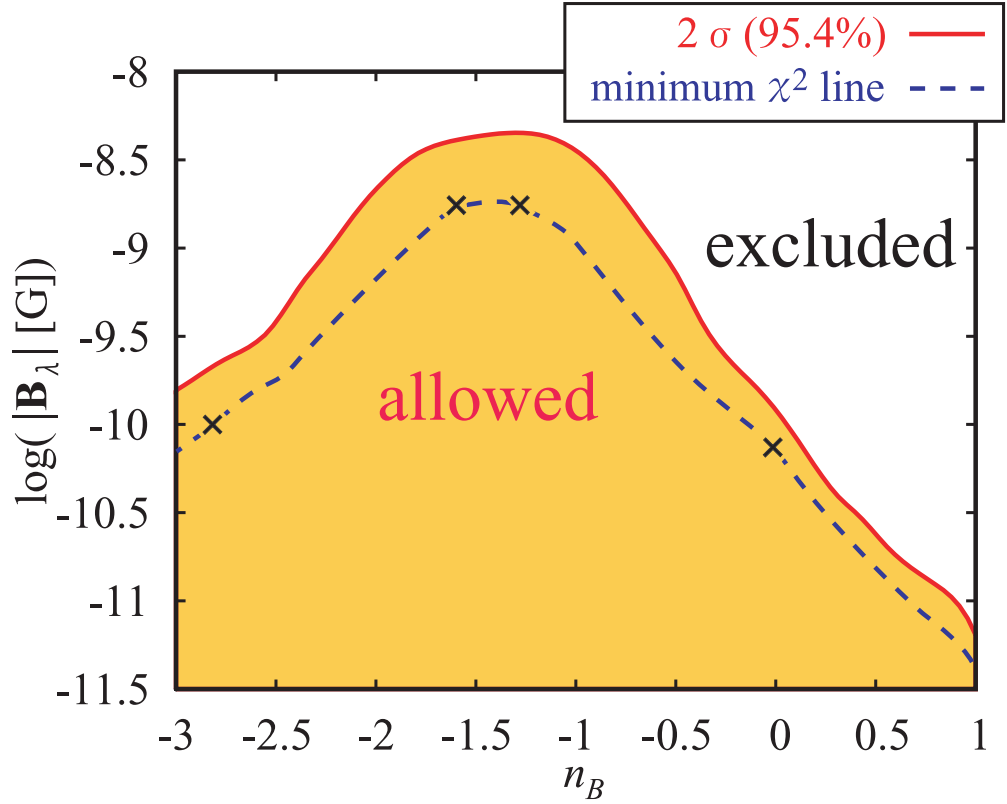}

\caption{Excluded and allowed regions at $2\sigma$(95.4\%) C.L.
 on two parameter plane $|\mathbf{B}_\lambda|$ vs. $n_B$, where $|\mathbf{B}_\lambda|$ is the primordial magnetic field strength and $n_B$ is the power-law spectral index. Thick dashed line is for a shallow minimum of the reduced $\chi^2\simeq$ 1.08, which runs through crosses ($\mathbf{\times}$) for ($|\mathbf{B}_\lambda|/$nG,$n_B$)=(0.1,-2.8), (1.9,-1.6), (1.6,-1.3), and (0.076,0.1).\label{fig1}}

\end{figure}

 We show the result of our MCMC analysis with WMAP data in the two parameter plane $|\mathbf{B}_{\lambda}|$ vs. $n_B$ in Fig.\ref{fig1}. Although the higher $l$ data from CBI(Mason et al., 2003), ACBAR(Kuo et al., 2004), and others may constrain the magnetic field strongly, we did not include those data in the present analysis because they have too large error bars. Therefore, the result is most sensitive to the WMAP data points at 500$\le l \le$900. Shown in Fig.\ref{fig1} is the excluded region at 2$\sigma$(95.4\%) C.L as bounded by the thick solid curve, but we could not find the lower boundary of the allowed region at the same  2$\sigma$(95.4\%) C.L. 
Note that we find a very shallow minimum of the reduced $\chi^2 \simeq 1.08$ along the thick dashed line which is almost parallel to the upper 2$\sigma$ boundary displayed as thick solid line is Fig.\ref{fig1}.
We can thus obtain the strength of primordial magnetic field $|\mathbf{B}|_\lambda \lesssim$3.9nG(2$\sigma$) at 1Mpc
 for the power-law spectral index $n_B\sim1.1$. 
This upper limit is the most reliable and newest one to the primordial magnetic field because we consider all effects on the CMB anisotropies, i.e. the effect of ionization ratio, the SZ effect, and the both scalar and vector mode effects of the magnetic field, for the first time, in the present estimate of $|\mathbf{B}_\lambda|$ and $n_B$.
In our numerical estimate of the magnetic field parameters, we continued the MCMC analysis until the other cosmological parameters converge well to the values listed in table 1.
The inferred parameter values are not very different from those of Spergel et al. (2002), but the important fact is that we cannot find obvious degeneracies of magnetic field parameters with other cosmological parameters.
We understand this for the following reasons. 
Primordial magnetic field is constrained by the cutoff scale of damping Alfven wave as indicated by Eq.(\ref{eq:kd}) (Jedamzik et al. 1998; Mack et al. 2002).
Since the cutoff scale is small enough compared with the mulitpoles $l \ll$ 3000 for current available data which we are interested in, the manifestation of this effect is remarkably seen as a monotonically increasing contribution from both scalar and vector perturbations in the mulutipole region higher then $l\sim$500 as far as $|\mathbf{B}|_\lambda \lesssim$3.9nG.
This sensitivity of the CMB power spectrum to the primordial magnetic field differs completely from those to the other cosmological parameters, which helps resolve the degeneracies among them.

There is, however, strong degeneracy between magnetic field strength $|\mathbf{B}_\lambda|$ and power spectral index $n_B$.
At this moment, we can not resolve this strong degeneracy, because the WMAP data are constrained to lower $l\le$ 900. 
The CBI and ACBAR data are in an interesting higher multipole region, but the error bars are too larger to resolve the degeneracy.
Precise data for higher $l$, where the effect of the primordial magnetic field is especially strong, are highly desirable. 

Let us shortly discuss the consistency of our upper limit $|\mathbf{B}_\lambda|\lesssim 3.9nG(2\sigma)$ with the other observational constraint on the magnetic field of the cluster of galaxies. 
Many astronomical observations indicate that the magnetic field strength in the present day cluster of galaxies is $\sim (0.1-1)\mu$G. Assuming isotropic collapse for cluster formation, one can straightforwardly estimate the magnetic field strength about $(1-10)$nG at the epoch of LSS. 
If the field strength is extremely larger than this critical value $\sim (1-10)$nG, we need some damping process, and if it is smaller than $\sim (1-10)$nG, on the other hand, we need amplification process.
We do not know any viable physical process of damping the magnetic field, but there are several amplification processes proposed in literature.
 We found wide allowed region in the two parameter plane ($|\mathbf{B}_{\lambda}| , n$) in Fig.\ref{fig1} for the primordial magnetic field which best fits the current CMB data.
Their parameter values satisfy the observational constraint form the cluster of galaxies $|\mathbf{B}_{\lambda}|\sim (1-10)$nG at the epoch of LSS.

To summarize, we studied the CMB fluctuation power spectrum by taking account of the scalar and vector modes from primordial magnetic field and the SZ effect.
The likelihood analysis of the WMAP data indicates that the upper limit of magnetic field strength is $|\mathbf{B}_\lambda| \lesssim 3.9$ nG (2$\sigma$) at 1 Mpc for any power spectral indices $n_B$ and reasonable values of the other cosmological parameters.

\acknowledgments{ We acknowledge Drs. K. Saigo, H. Hanayama, M. Higa, R. Nakamura, K. Umezu, H. Ohno, K. Takahashi, M. Oguri, Profs. M. Yahiro and G. J. Mathews for their valuable discussions. This work has been supported in part by Grants-in-Aid for Scientific
Research (13640313, 14540271) and for Specially Promoted Research (13002001)
of the Ministry of Education, Science, Sports and Culture of Japan, and the
Mitsubishi Foundation. K. I. also acknowledges the support by Grant-in-Aid for
JSPS Fellows.
}


\begin{deluxetable}{llll}
\tablecolumns{4}
\tablewidth{0pc}
\tablecaption{Calculated $\Lambda$CDM model parameters which best fits the
WMAP data with the effect of primordial magnetic field taken into account.
}
\tablehead{
Parameter & Mean and 68\% C.L. Errors &  95\% C.L. Errors
}
\startdata
$\Omega_b h^2$ & $0.023_{-0.001}^{+0.001}$ & $_{-0.002}^{+0.002}$ \\
$\Omega_{CDM} h^2$ & $0.13_{-0.01}^{+0.02}$ & $_{-0.03}^{+0.03}$ \\
$n_s$ & $1.00_{-0.03}^{+0.03}$ & $_{-0.05}^{+0.05}$ \\
$A_s$ & $0.87_{-0.14}^{+0.14}$ & $_{-0.20}^{+0.20}$ \\
$\tau$ & $0.11_{-0.04}^{+0.04}$ & $_{-0.09}^{+0.10}$ \\
$h$ & $0.70_{-0.04}^{+0.04}$ & $_{-0.09}^{+0.08}$ \\
\enddata
\tablecomments{
Calculated corresponding cosmic expansion ages are $t_0$/Gyr = $13.3_{-0.3}^{+0.3}$(68\% C.L.) and $13.3_{-0.5}^{+0.5}$(95\% C.L.).}
\end{deluxetable}

\end{document}